\documentclass[final,twocolumn,prl,aps,floats,amsfonts,amssymb,showpacs,superscriptaddress]{revtex4}
\usepackage{graphicx}
\usepackage{color}
\usepackage{psfrag}
\usepackage{gensymb}
\usepackage{enumerate}
\usepackage{latexsym}
\usepackage{amssymb}
\usepackage{amsmath}
\usepackage{wasysym}
\usepackage{hyperref}
\usepackage{color}

\begin{document}

\title{Dual local and non-local cascades in 3D turbulent Beltrami flows}

\author{E. Herbert, F. Daviaud, B. Dubrulle}
\affiliation{SPHYNX, Service de Physique de l'\'Etat Condens\'e, DSM, CEA 
Saclay, CNRS URA 2464,
91191 Gif-sur-Yvette, France}
\author{S. Nazarenko}
\affiliation{Mathematics Institute, University of Warwick, Coventry CV4 7AL, UK }

\author{A. Naso}
\affiliation{Laboratoire de Mécanique des Fluides et d’Acoustique, CNRS UMR 5509, École Centrale de Lyon, Université de Lyon, France}

\date{\today}


\begin{abstract}

We discuss the possibility of dual local and non-local cascades in a 3D turbulent Beltrami flow, with inverse energy cascade and direct helicity cascade, by analogy with 2D turbulence. We discuss the corresponding energy spectrum in both local and non-local case. Comparison with a high Reynolds number turbulent von Karman flow is provided and discussed.

\end{abstract}

\maketitle

Turbulent flows are usually characterized by a wide range of scales of motion. Despite their apparent randomness, these motions organize themselves in a hierarchical manner, resulting in power-law spectra for the energy as a function of the wavenumber $k$. The index of the power law and its range of scale depends on the turbulence geometry. For example, in forced isotropic incompressible turbulence, the energy spectrum varies like $k^{-5/3}$ at large scale in 2D and in the so-called  inertial  range in 3D, while it varies like $k^{-3}$ at small scales in 2D \cite{kraichnan67}. These features are traditionnally explained through local cascades of inviscid global quadratic invariants of the Navier-Stokes equations. In 2D, the conserved quantities are the kinetic energy $E=(1/2)<{\bf v}^2>$ and the enstrophy $\Omega=<{\bf \omega}^2>$ where ${\bf\omega}=\nabla\times {\bf v}$ is the vorticity and $v$ the velocity. In 3D, the two quadratic invariants are the kinetic energy $E$, and the helicity $H=<{\bf \omega}\cdot{\bf v}>$, that is sign-indefinite, in contrast to energy and enstrophy that are both non-negative. This difference has been originally recognized as responsible for the difference between 2D and 3D situation \cite{brissaud73,kraichnan73} : in 2D, the enstrophy invariant effectively blocks the energy cascade from large to small scales, resulting in a dual cascade, with energy cascading to large scales, and enstrophy to small scales \cite{kraichnan67}. In 3D isotropic incompressible turbulence, a joint helicity and energy cascade is rather expected \cite{kraichnan73,borue97,ditlevsen01}, in which both the energy and the helicity cascade towards the small scales. A possible mechanism allowing coexistence of constant energy and helicity flux is the near cancellation of the helicity flux between modes of different helical polarity \cite{chen02}.\

Once the cascade scenario has been clarified, the computation of the index of the energy spectrum follows by simple dimensional argument on the cascade transfer rate. For example, in 3D, if one assumes that the cascade is local and determined only by the energy flux $\epsilon$, one gets an energy spectrum $E(k)\sim \epsilon^{2/3} k^{-5/3}$ and a helicity spectrum $H(k)\sim \delta \epsilon^{-1/3} k^{-5/3}$, where $\delta$ is the helicity flux. In 2D, similar assumptions using the enstrophy flux $\eta$ instead (when there is no energy flux) leads to $E(k)\sim \eta^{2/3} k^{-3}$ (with logarithmic corrections that may result in a steeper spectrum) and an enstrophy spectrum $\Omega(k)\sim \eta^{2/3}  k^{-1}$ at the small scales. In contrast, if one assumes a non-local cascade, as suggested in \cite{laval99} at a rate determined by the large scale shear rate $S$, one gets the same spectrum but with a different prefactor : $E(k)\sim (\eta/S) k^{-3}$ and $\Omega(k)\sim (\eta/S) k^{-1}$ \cite{nazarenko00}. Numerical simulations rather supports this second scenario \cite{laval99}.\

In the present paper, we consider an intermediate situation between the 2D and 3D geometries, in which the basic flow is Beltrami, so that vorticity and velocity are colinear everywhere and helicity is maximal. In this case, the helicity is of constant sign and can possibly block the forward energy cascade, resulting in a dual cascade of energy and helicity. We then consider both local and non-local cases, and compute the resulting energy and helicity spectrum. We argue that such a special situation is empirically encoutered in large Reynolds number von Karman turbulent flows and compare our predictions with energy spectra obtained in these flows.

\begin{figure}[h]
\begin{minipage}{\columnwidth} 
  \includegraphics[width=\textwidth]{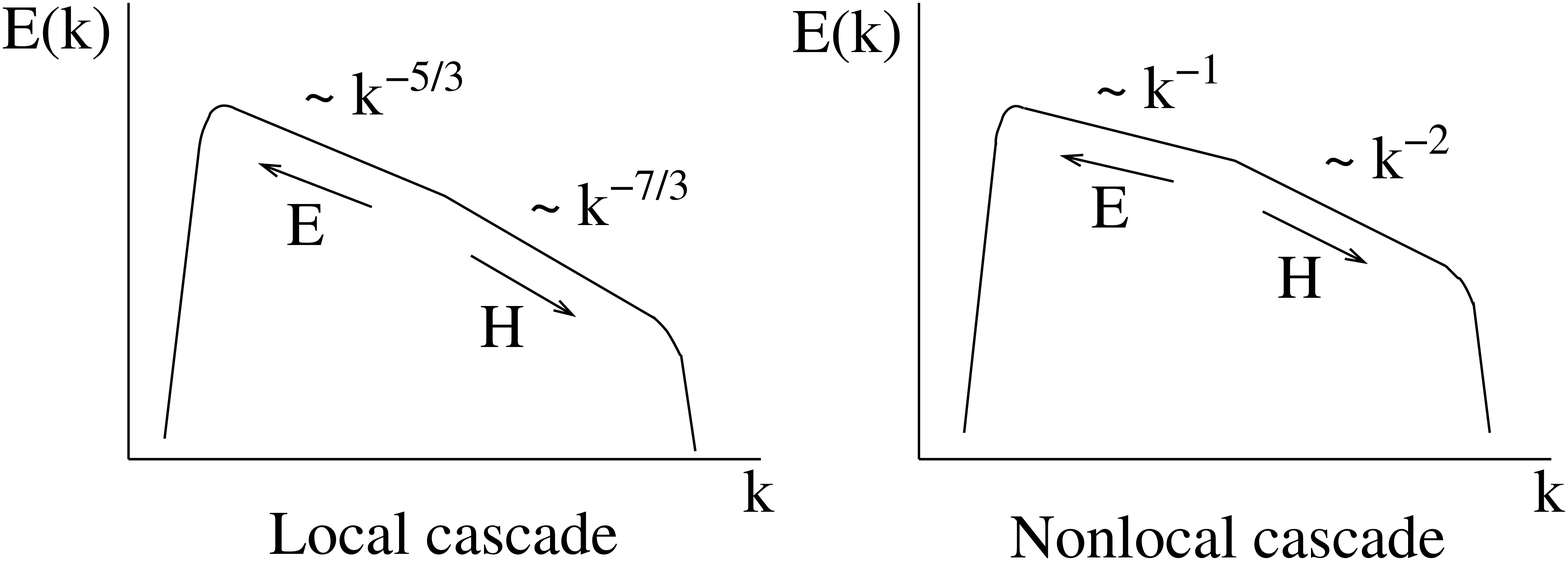}
\end{minipage}
\caption{Summary of the two different possibilities for dual helicity and energy cascades as a function of the wavenumber $k$ in a Beltrami flow. Left : local case; right : non-local case.}
\label{sigmaofxi}
\end{figure}

\paragraph*{Theoretical background and definitions}

We consider a 3D Beltrami flow, such that ${\bf \omega}=\mu {\bf v}$, where $\mu$ is constant. This flow is an exact solution of 3D Euler equations. It also describes the stationary state of a high Reynolds number von Karman flow \cite{monchaux06}. In such a flow, vorticity is maximal, helicity $H=\int {\bf v}\cdot {\bf\omega}=\int dk H(k)$ is of constant sign and we have the equality $H(k)=k E(k)$ at the wave number of the Beltrami flow (while in general, we have only the inequality $H\le k E$). This relation is the analog of the relation $\Omega=k^2 E$ in 2D turbulence. It has some important consequences onto the possible cascade direction. Indeed, if there exists a non-zero flux of energy at small scale $k>1$, it will be smaller than the corresponding helicity flux by a factor $1/k$. Therefore, at the small scales, the cascade will be mostly governed by the helicity transfer rate $\delta$ and will correspond to a direct helicity cascade. At the large scales $k<1$, the energy flux dominates instead, and we shall have an inverse energy cascade. To compute the corresponding spectra, we introduce
the rate of energy transfer $\Pi_E\cong\frac{kE(k)}{\tau(k)}$ and the rate of helicity transfer $\Pi_H\cong\frac{kH(k)}{\tau(k)}$, where $\tau(k)$ is a characteristic transfer time at wavenumber $k$. We consider below two situations. In the {\sl local} case, $\tau$ is governed only by interaction with wave numbers that are of comparable size. The only characteristic time scale one can build in such a case is an  eddy turn-over time, given on dimensional ground by
\begin{equation}
\tau_{loc}(k)=\left(\int_0^k q^2 E(q) dq\right)^{-1/2}.
\label{localtime}
\end{equation}
In the {\sl non-local} case, transfers between scales are dominated by interactions involving two well separated scales. This situation is automatically realized for spectra steeper than $k^{-3}$ since in that case, the integral defining $\tau_{loc}$ is dominated by the behaviour near the minimum wavenumber, instead of by the behaviour near $k$. Non-local transfers also arise in situations with a well defined large scale  mean flow, that produces both sweeping and shearing of small scales over a characteristic time scale $S^{-1}$, where $S$ is the large scale shear rate. In such a case, the transfer time is given by the time scale of the largest scale. This is an intrinsic property of the flow which applies to both Energy and Helicity resulting in 
\begin{equation}
\tau_{non-loc}=S^{-1}.
\label{nonlocaltime}
\end{equation}

We may now proceed to the spectra derivation. Consider first the small scale energy transfer, for $k>1$. In such a case, the helicity transfer is dominant. Assuming furthermore the existence of an  inertial  range of scale, in which this helicity transfer is equal to a constant $\delta$, we get :
\begin{equation}
\Pi_H=\frac{kH}{\tau(k)}=\delta.
\label{smallscale}
\end{equation}
In the local case, this leads to the integro-differential equation :
\begin{equation}
k^2 E(k)\left(\int_0^k q^2 E(q) dq\right)^{1/2}=\delta,
\label{integross}
\end{equation}
with solution
\begin{eqnarray}
E^{loc}(k) &\sim& \delta^{2/3} k^{-7/3},\\
&& \rm{(local\; direct\; helicity\; cascade)}.\nonumber
\label{integrosssol}
\end{eqnarray}
In the non-local case, we get directly~\cite{nazarenko00} :
\begin{eqnarray}
E^{non-loc}(k) &\sim& \delta S^{-1} k^{-2},\\
&& \rm{(non-local\quad direct\quad helicity\quad cascade)}.\nonumber
\label{integrosssolnl}
\end{eqnarray}
Consider now the large scale transfer. It is dominated by energy transfer. Assuming furthermore an inertial range of scales with constant energy transfer rate $\epsilon$, we then now get :
\begin{equation}
\Pi_E=\frac{kE}{\tau(k)}=\epsilon.
\label{smallscaleE}
\end{equation}
In the local case, this now leads to the integro-differential equation :
\begin{equation}
k E(k)\left(\int_0^k q^2 E(q) dq\right)^{1/2}=\epsilon,
\label{integrossE}
\end{equation}
with solution
\begin{eqnarray}
E^{loc}(k) &\sim& \epsilon^{2/3} k^{-5/3},\\
&& \rm{(local\; inverse\; energy\; cascade)}.\nonumber
\label{integrosssolE}
\end{eqnarray}
In the non-local case, we get directly :
\begin{eqnarray}
E^{non-loc}(k) &\sim& \epsilon S^{-1} k^{-1},\\ 
&&\rm{(non-local\; inverse\; energy\; cascade)}.\nonumber
\label{integrosssolnlE}
\end{eqnarray}
These results are summarized in Fig. 1. We see that the difference between local and non-local spectrum can be tested both via the slope of the spectra, and using the prefactor of the spectra, at variance with the 2D situation where at small scale, both local and non-local hypotheses lead to the same $k^{-3}$ spectrum, with different prefactor.

We now argue that these new energy spectra, resulting from a dual energy and helicity cascade, are indeed observable in real flows. For this, we consider a turbulent von Karman flow, where it has been empirically shown that the stationary state is a Beltrami state \cite{monchaux06,monchaux08}.


\paragraph*{Experimental setup}

We have worked with a specific "axisymmetric" configuration: the von 
K\'arm\'an flow generated by two counter-rotating impellers in a 
cylindrical vessel \cite{cortet_susceptibility_2011}. The cylinder radius and height are respectively 
$R=100$\,mm and $H=180$\,mm. The impellers consist of $185$\,mm diameter 
disks fitted with sixteen $20$\,mm high curved blades. The impellers 
are driven by two independent motors at frequencies $f_1 = f_2$. 
The mean forcing frequency is $f=(f_1+f_2)/2$. The flow is divided into two toric cells separated by an azimuthal shear layer. In the exact counter-rotating regime, in a statistical sens the shear layer is located at mid-distance of each impeller.

In the present paper, we force the flow through disk rotation in the (+) direction associated with the  convex face of the blades going forward \cite{cortet_susceptibility_2011}. The working fluid is either water or glycerol 
at different dilution rates (respectively 100\% glycerol, 74\% glycerol and pure water). The resulting accessible Reynolds numbers ($Re=2\pi f R^2 \nu^{-1}$ with $\nu$ the kinematic viscosity) vary from 
$10^2$ to $10^6$. 

Measurements are done thanks to a stereoscopic Particle Image Velocimetry (S-PIV) system.
This system provides time series of the 3 components of the velocity (radial $v_r(r,z,t)$, vertical $v_z(r,z,t)$ and azimuthal $v_\phi(r,z,t)$), and of the azimuthal vorticity component $\omega_\phi=\partial_z v_r-\partial_r v_z$ in the meridian plane on a $58\times 63$ points grid. Time series sampling frequency is in the range  1.7 to 15\,Hz and includes 2400 to 4200 samples. Fluctuations of the velocity and vorticity are then obtained after substraction of the mean velocity field to the instantaneous velocity field.  From the corresponding time series, we compute the 3 dimensions space-time power spectrum $E(\boldsymbol{k},2\pi f)$, where $\boldsymbol{k}$ is the wavevector and $f$ is the frequency. We deduced the spatial velocity spectrum $E(k=||\boldsymbol{k}||)$ (or equivalently $E(1/\lambda)$, where $\lambda= 1/||\boldsymbol{k}||$ is the reduced wavelength)  by integrating over all directions of $\boldsymbol{k}$ and over all frequencies. Note that our procedure defines genuine spatial spectra, independently of any Taylor hypothesis.


\paragraph*{Results}

\begin{figure}[h]
  \begin{minipage}{\columnwidth} 
    \includegraphics[width=0.49\textwidth]{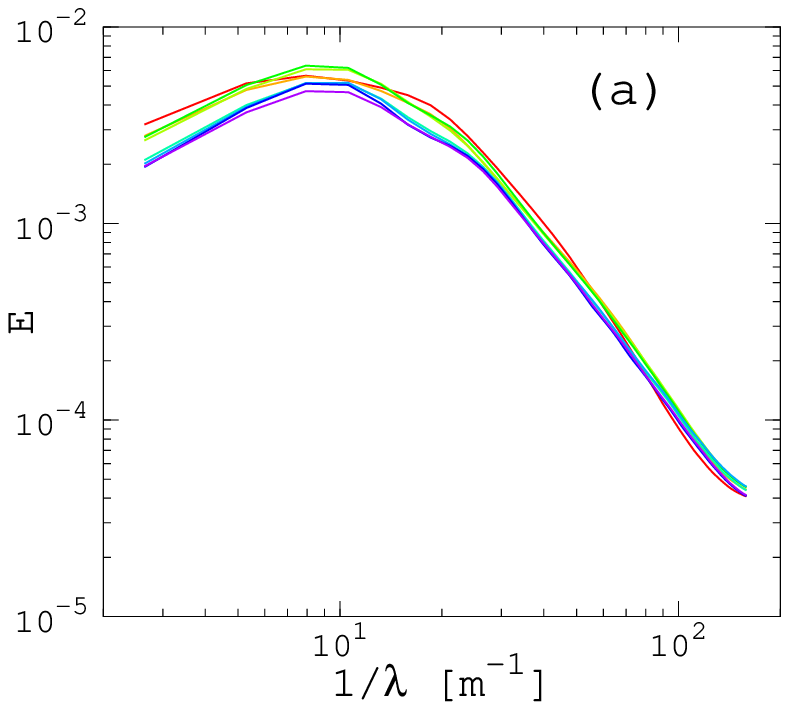}
    \includegraphics[width=0.49\textwidth]{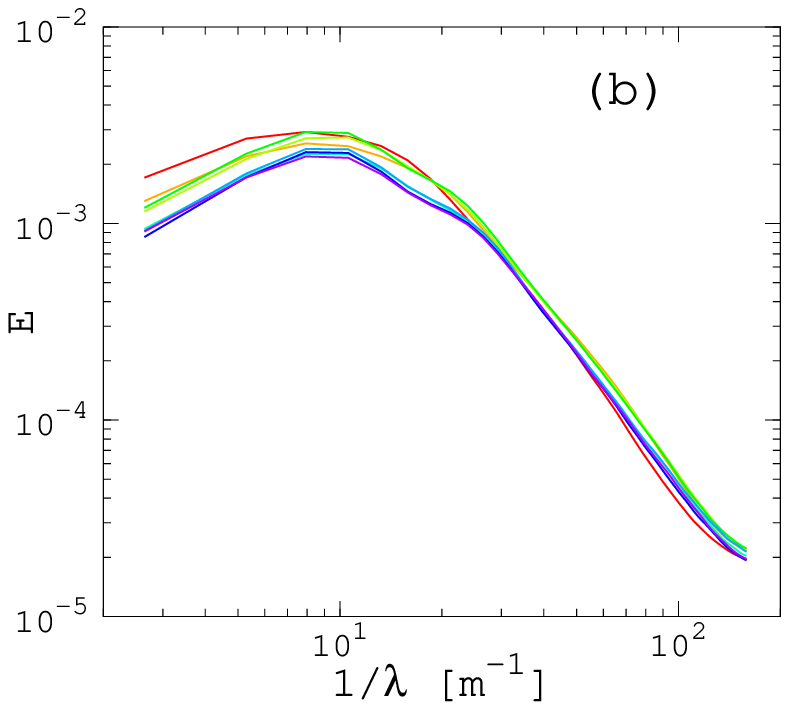}
  \end{minipage}
  \begin{minipage}{\columnwidth} 
    \includegraphics[width=0.49\textwidth]{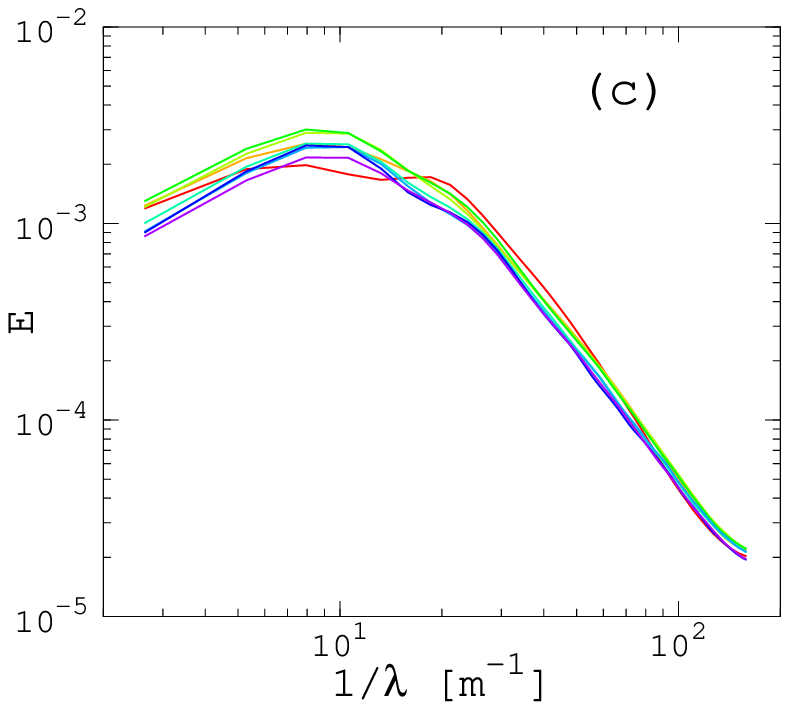}
     \includegraphics[width=0.49\textwidth]{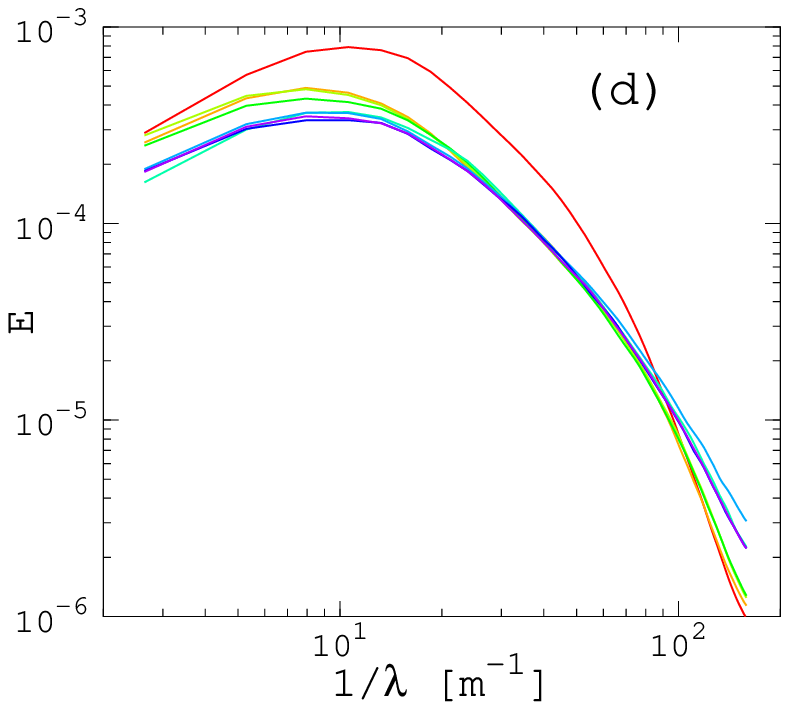}
  \end{minipage}
\caption{Spatial spectrum $E(1/\lambda)$ in arbitrary units of the velocity fluctuations at Reynolds number $\mathrm{Re}/1000=3.2$ (red), 9.5 (yellow), 21 (light green), 41(green), 122 (emerald), 266 (light blue), 410 (dark blue) and 1022 (purple). (Color online): a) Total; b) $r$-component; c) $\phi$-component; d) $z$-component. The energy spectra have been normalized by $R^2f^2$.}
\label{SpeAllRe}
\end{figure}
The spatial energy spectra for the three components of velocity at different Reynolds numbers are shown in Fig.~\ref{SpeAllRe}. In all cases, the energy is mainly concentrated at the largest scales, around $\lambda=0.1$ m, corresponding to the cylinder radius. In the azimuthal and radial components, as well as in the total energy,  one observes a well defined inertial range at small scales $\lambda<0.03$ m, and the very beginning of another scaling range at scales between $0.03<\lambda<0.1$ m for large enough Reynolds numbers.   However, no clear scaling for the spectra is observed in the vertical component.  This is a classical outcome in anisotropic turbulence (see e.g. \cite{minini2010}) because of different scaling in the directions parallel and perpendicular to the anisotropy. In our case, the rotation of the impellers is along the $z$-direction, defining the natural direction of anisotropy. Moreover, we have checked that most of the energy is contained in the azimuthal velocity component, i.e. in the direction perpendicular to the rotation axis, in agreement with recent numerical simulations of helical turbulence \cite{minini2010}. In the sequel, we therefore focus on the azimuthal velocity $v'_\phi$ and vorticity $\omega'_\phi$ fluctuations, and on the corresponding energy $E_\phi=<v_\phi^{'2}>=\int E(k)dk$ and helicity $H=<v'_\phi \omega'_\phi>=\int H(k)dk$.

\begin{figure}[h]

\begin{minipage}{\columnwidth} 
  \includegraphics[width=0.9\textwidth]{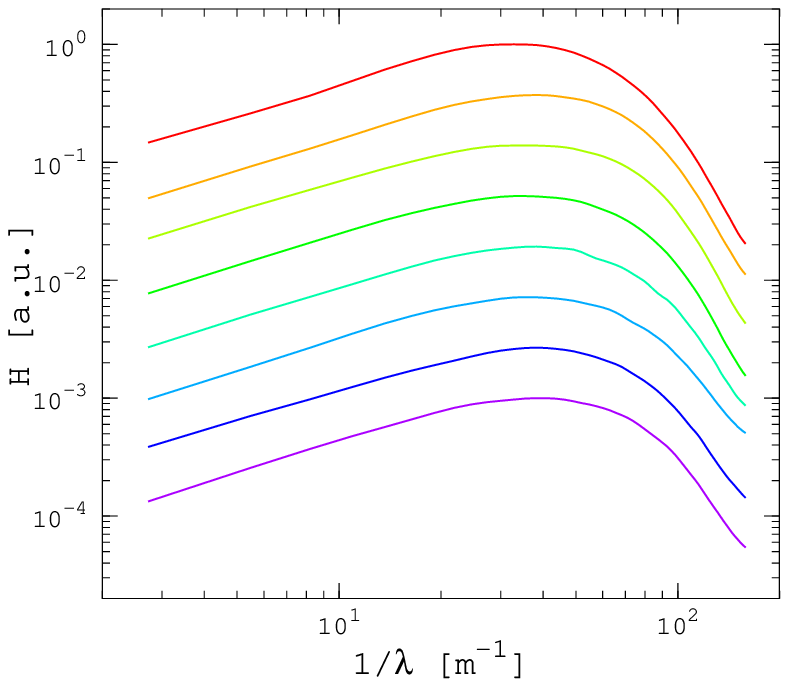}
\end{minipage}

\begin{minipage}{\columnwidth} 
 \includegraphics[width=0.9\textwidth]{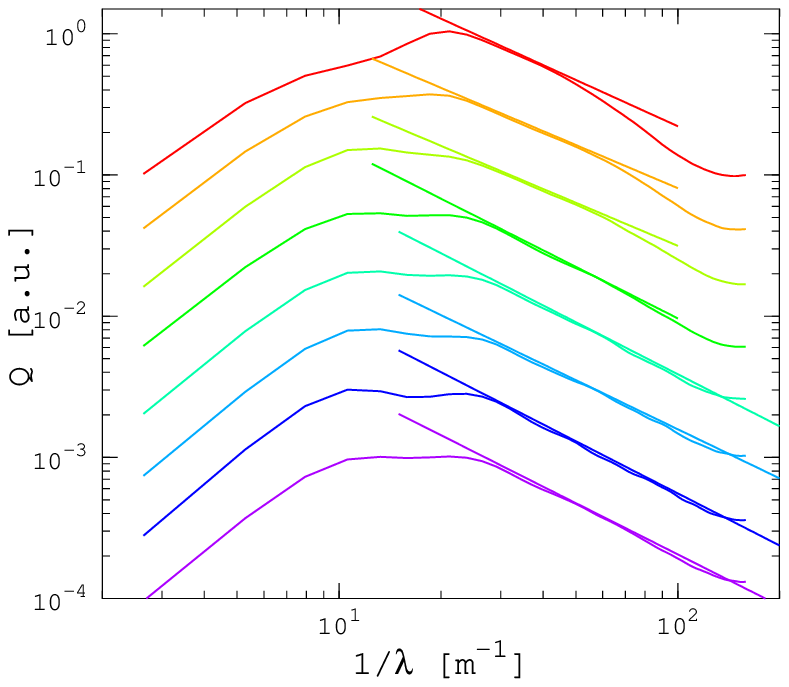}
\end{minipage}

\caption{Top: azimuthal contribution of helicity  $H=\int v_\phi \omega_\phi$. Bottom: spatial spectrum $Q=E_\phi(1/\lambda)/\lambda$ of the azimuthal velocity fluctuations. The spectra have been arbitrarily shifted for clarity. Same color code as in Fig.~\ref{SpeAllRe}. Lines are linear fittings in the direct cascade range $25<\lambda^{-1}<50$ (resp. $30<\lambda^{-1}<100$) when $Re\le 41 000$ (resp. $Re \ge 65 000$).}
\label{compaSpecRe}
\end{figure}

As a first preliminary check, we compare  in Fig.~\ref{compaSpecRe} the azimuthal helicity spectrum, with $E_\phi(1/\lambda)/\lambda$. If the flow is indeed Beltrami, the two spectra should coincide  over the range of scale defining the Beltrami flow. We see that they indeed look similar at large scale, which is consistent with the fact that the mean (large scale) velocity field in von Karman is a Beltrami flow \cite{monchaux06}. A more quantitative test of the Beltrami property would however require the three components of vorticity, which is not available in our experimental set-up.

To get a description of the possible cascades, we now concentrate on the energy spectra behaviors, through the compensated spectrum  $Q=E_\phi(1/\lambda)/\lambda$.
At the smallest Reynolds number (red curve), $Q$ is maximal  at $\lambda_c\approx 1/25$\,m corresponding to the forcing scale. As the Reynolds number is increased, we see that for high enough Reynolds numbers (Re$>\sim 40.10^3$) $Q$ is developping at large scale a constant range starting at $\lambda_c$ and extending progressively towards the size of the vessel $\lambda\approx 0.1 $m.  
This corresponds to a range of $k^{-1}$ scaling for the energy spectrum, in  agreement with the scaling expected for the inverse cascade in the non-local case.  
Below $\lambda_c$, a crossover for $Q$ is observed at any Reynolds number.  This corresponds to the transition to the direct cascade regime where a power law can be measured. We observe that spectra at small scales are gradually filled with increasing Reynolds number from $\lambda \approx 1/50 $\,m up to reach the resolution limit at $\lambda\approx 1/130 $\,m.

\begin{figure}[h]

\begin{minipage}{\columnwidth} 
\includegraphics[width=0.8\textwidth]{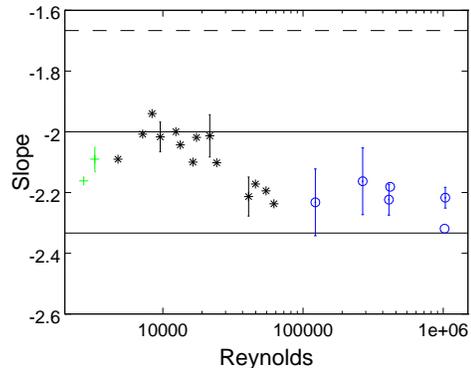} 
\end{minipage} 

\caption{Slopes of $E(k)$ for various Reynolds numbers (color online). Green crosses, black stars and blue circles  corresponds to resp. 100\%, 74\% glycerol mixture and pure water. Error bars are computed using the same time series as in Fig.~\ref{compaSpecRe}. Horizontal black lines mark the  local (-7/3) and non-local (-2) predictions. The dashed line marks the Kolmogorov (-5/3) index.}
\label{Slope}
\end{figure}

The  scaling slope in the direct cascade regime is computed using linear fit with adaptative boundaries. The resulting  slopes  are shown in Fig.~\ref{Slope}  for a large range of Reynolds number. 

The first striking result is that for any Reynolds number, the direct cascade slope is steeper than the  classical Kolmogorov -5/3 expected in the homogeneous and isotropic 3D turbulence. 
At small Reynolds number ($\mathrm{Re} \leq 50.10^3$), the slope is nearly  $ -2$ close to the non-local prediction. The slope then steepens further at increasing Reynolds number  reaching $ -2.2$ at $\mathrm{Re} \geq 100.10^3$, close to the local prediction (-7/3). At the present time, our time statistics (3600 frames) does not allow us to restrict the error bars on the fit below 10 percent. This is enough to exclude the Kolmogorov case (exponent $-5/3$) from our results, but not to assess without ambiguity the observed shift from a "non-local" to a "local" spectrum at $Re \simeq 5.10^4$. It is however interesting to note that the transition Reynolds number is precisely associated with a spontaneous symmetry breaking bifurcation of the mean flow \cite{cortet_susceptibility_2011}, with possible associated changes in the structure of the vortices of the shear layer. Work is in progress to check whether this is likely to change the mean shear felt by the smallest scale, and, therefore, the locality properties of the cascade.


\paragraph*{Discussion}

We have studied spatial energy spectra in a von Karman experiment, in a wide range of Reynolds numbers. In all cases, we have observed a dual energy cascade, one at the scales larger than the forcing scale, with increasing energy concentration at increasing Reynolds number, and one at the scales smaller than the forcing scale, with increasingly steeper slope as the Reynolds number increases. None of these cascades scale according to the $k^{-5/3}$ Kolmogorov prediction, valid for isotropic homogeneous turbulence. On the other hand, their slope is in agreement with the slopes predicted using the helicity conservation in a Beltrami like flow. Given that the von Karman experiment is neither isotropic, nor homogeneous, the discrepancy with Kolmogorov prediction is not surprising. In addition, its Beltrami character at large scale \cite{monchaux06} makes it suitable to test the theoretical prediction. It is nevertheless interesting to compare our experimental findings with other non-isotropic situations, where non-Kolmogorov spectra have been predicted and observed. For example, in the rotating helical case, Mininni and Pouquet \cite{minini2010} observe a similar coexistence of a large scale inverse energy cascade, and a direct cascade towards small scales. The slope of the energy and helicity spectra in the direct cascade is close to $-2$, like in the present experiment. However, using argument adapted from Iroshnikov-Kraichnan
\cite{TOTO}, they predict that for flow with maximal helicity, the scaling should be $k^{-5/2}$ for the energy, and $k^{-3/2}$ for the helicity. This is not observed in our experiment, although we are in a situation with large (but perhaps nor completely maximum) helicity.
It would be interesting in that respect to perform new von Karman experiments with added rotation, to see whether a steepening of the energy spectrum occurs.

{\bf Acknowledgments}\
We thank the CEA programm DSM-Energie for support, A. Chiffaudel for fuitful discussions, and P-P. Cortet for the PIV data.

\end{document}